\documentclass[10pt,aps,pra,twocolumn,showpacs,superscriptaddress]{revtex4-2}
\usepackage[english]{babel} 	
\usepackage[usenames, dvipsnames]{color} %color management
\usepackage{graphicx} % enhanced support for graphics
\usepackage{bm} % Bold mathsymbols
\usepackage{amsmath} % improvements for mathformulas
\usepackage{amsthm} % improvements theorem-like enviouronment
\usepackage{amssymb} % extended symbols
\usepackage{times} % fontes adobe
\usepackage[pdftex,colorlinks=true,urlcolor= blue,linkcolor=Blue,citecolor=RedViolet]{hyperref}
\usepackage{multirow}
\usepackage{graphicx}
\usepackage{bm}% bold math
\usepackage{physics}
\usepackage{ bbold }

\providecommand{\bra}[1]{\langle #1 |}
\providecommand{\ket}[1]{| #1 \rangle}

\providecommand{\ketbra}[2]{|  #1\rangle \langle #2 |}

\usepackage[small,bf]{caption}

\usepackage{soul}

\begin{document}

\title{Variational-quantum-eigensolver-inspired optimization for spin-chain work extraction}

\author{Ivan Medina}
\email{ivan.medina@ifsc.usp.br}
\author{Alexandre Drinko}
\author{Guilherme I. Correr}
\author{Pedro C. Azado}
\author{Diogo O. Soares-Pinto}
\affiliation{Instituto de F\'{i}sica de S\~{a}o Carlos, Universidade de São Paulo, CP 369, 13560-970 São Carlos, Brazil}

\begin{abstract}
The energy extraction from quantum sources is a key task to develop new quantum devices such as quantum batteries (QB). In this context, one of the main figures of merit is the ergotropy, which measures the maximal amount of energy (as work)  that can be extracted from the quantum source by means of unitary operations. One of the main issues to fully extract energy from the quantum source is the assumption that any unitary operation can be done on the system. This assumption, in general, fails in practice since the operations that can be done are limited and depend on the quantum hardware (experimental platform) one has available. In this work, we propose an approach to optimize the extractable energy inspired by the variational quantum eigensolver (VQE) algorithm. In this approach, we explicitly take into account a limited set of unitaries by using the hardware efficient asatz (HEA) class of parameterized quantum circuits. As a QB we use an one-dimensional spin chain described by a family of paradigmatic first neighbor Hamiltonians such as the $XXX$,$XXZ$, $XYZ$, $XX$, $XY$ and transverse Ising models. By building our parameterized quantum circuits assuming that different types of connectivity may be available depending on the quantum hardware, we numerically compare the efficiency of work extraction for each model. Our results show that the best efficiency is generally obtained with quantum circuits that have connectivity between first neighbor spins.
\end{abstract}

\maketitle

\section{Introduction}

The development of new techniques and protocols to efficiently store, transfer, and use energy on demand is arguably one of the most important goals of human kind currently \cite{Dubi,AlexiaQEI}. In the last decades, the emergent fields of quantum information and quantum thermodynamics allowed us to explore how intrinsic quantum phenomena can impact the energetics of quantum systems \cite{Binder2018,PhysRevA.99.062103,micadei2019reversing,ivan3,IvanHenao,Andres}. In this context, Alick and Fannes proposed \cite{Alicki2013} a device termed the quantum battery (QB), in which they sought to explore how non-classical features such as entanglement and coherence could be used as resources to efficiently extract work from an ensemble of quantum systems. After this seminal work, QBs have been extensively explored with the perspective that these resources could be used  to achieve better performance than their classical counterparts \cite{PhysRevLett.125.180603,tirone2023quantum,Binder_2015,PhysRevB.104.245418,PhysRevE.102.042111,PhysRevLett.131.060402}. For instance, a very promising property that arose from QBs composed by multiple quantum systems is the possibility to have a superextensive power charging scaling, which means that charging power grows faster than the number of systems in the QB \cite{MicrocavityQB}.

When dealing with QBs, one of the central figures of merit is the ergotropy \cite{ergotropy2,ergotropy,Alicki2013,sone2021quantum}. The ergotropy quantifies the maximum work that can be extracted by means of unitary operations, i.e.,  without changing the QB's entropy \cite{sone2021quantum}. Naturally, the ergotropy  depends on the quantum state of the QB (or quantum source) and on its Hamiltonian. In the last years, much effort has been made to find optimal processes to charge and discharge the QB in a variety of theoretical models \cite{optimal,optimal2,optimal3,optimal4,optimal5,optent,VQErg,PhysRevLett.130.210401,PhysRevLett.120.117702,lossfreeEx,catalano2023frustrating,tirone2022quantum}. This includes, for example, proposes of many-body QBs in which their Hamiltonians have interactions between subsystems and, consequently, could have entangled eingenspectrum. Models of this type are known as spin-chain QBs. The first spin-chain QB model was proposed by Le {\it et al.} \cite{Tao}, where they studied the role of anisotropy in the energetics of the Heisenberg and Ising spin chain models. Further works showed how disorder in the interactions and the dimensionality of the spin chain can enhance the performance of the QB \cite{spinchainbattery1,spinchainbattery2}. The optimal charging of dissipative spin-chain batteries was also explored in Ref. \cite{optimal5}, where  feedback control techniques were used. Another interesting result regarding spin chain models for many-body QBs is that working near phase transitions points can enhance the extractable energy and the thermodynamic efficiency \cite{batteryphase}. Despite all the theoretical results,  there are still just a few experimental works dealing with QBs in general \cite{MicrocavityQB,expAlexa,expCQE,expMahesh,Zheng_2022}, and even for proof of principles experiments there are still challenges to overcome. One of the difficulties is that, in practice, optimizing the extractable work is a hard task because the unitary transformations that can be done are limited and depend on the experimental platform. Strategies to optimize the extractable work in realistic scenarios are open problems. A very recent review on the advances and the state of art in QBs can be found in Ref. \cite{batteryreview}.

% Many effort has been done in the last years to find optimal process to charge and discharge the QB in a variety theoretical models \cite{}. 
%Bounds has been found to maximum work extraction and optimal charging power considering classical and non-classical effects\cite{}. 
%However, very few works deal with realistic implementations of quantum batteries up this date \cite{}. One of the main problems is the complexity of implement entangling operations (in order to harness quantum effects on the QB protocol) in the case of QBs composed by a large number of cells which have an internal interacting hamiltonian. A very recent review on the advances of QBs can be found in \cite{batteryreview}.

At the same time, the growth in complexity of quantum computational tasks to address realistic problems in noisy intermediate scale quantum (NISQ) devices demands the development of novel optimization protocols \cite{preskill2018quantum}. We can mention, among others, portfolio optimization \cite{canabarro2022quantum,brandhofer2022benchmarking,yalovetzky2021nisq,kubo2023pricing}, quantum machine learning \cite{biamonte2017quantum, schatzki2022theoretical, PRXQuantum.4.010328,benedetti2019parameterized,schuld2015introduction,cerezo2022challenges}, and the simulation of molecules, materials, and other many-body systems \cite{delgado2021variational,lee2021neural,stetcu2022variational,PhysRevB.106.214429}. A very exciting way to deal with such complex problems is using variational quantum algorithms (VQA), which have proved to be a powerful tool in a multitude of different contexts \cite{Cerezo2021VqaReview}. In particular, the VQA named the variational quantum eigensolver (VQE) \cite{peruzzo2014variational,tilly2022variational} is specifically designed to minimize the energy of many-body systems and find the ground state of their Hamiltonians. Inspired by these remarkable results, the goal of this work is to propose a VQE-inspired optimization to address the optimization of the work extraction from spin chain QBs. To do this, we choose a broad family of spin chain Hamiltonians as a QB. Moreover, we consider a realistic limited set of quantum operations to perform the work extraction and also that we may have different connectivity depending on the quantum hardware one has available. With this, we expect that our protocol can be applied to any experimental platform that is currently available. It is worthwhile to mention that some recent works used machine learning techniques to optimize the charging and discharging processes of quantum batteries \cite{rfbattery,aibattery,VQErg}. In particular, in Ref. \cite{VQErg} a new scheme called variational quantum ergotropy (VQErgo) was proposed to estimate the ergotropy in quantum batteries.

%The idea is to use variational quantum algorithms, which are on the spotlight, as an effective and experimentally way to optimize QB work extraction. In principle, we can focus in the Hardware Efficient Ansatz (HEA), used in the algorithm called variational quantum eigensolver (VQE). Mainly because this ansatz uses shallow circuits that can effectively explore large Hilbert spaces of many body systems. In this way, many of the circuits is expected to be implementable in the majority of the quantum platforms we have available. 

%The use of this hamiltonian it is not simply illustrative. In the context of quantum batteries, it has been shown that spin-spin  interactions lead can yield an advantage in the charge power if compared to non interacting cases \cite{tao}. 

This paper is organized as follows. We start by introducing the main tools of our work in Sec. \ref{pre}. In Sec. \ref{Work extraction}, we briefly review the concept of work extraction and ergotropy. In Sec. \ref{spinchain} we introduce the spin chain Hamiltonians describing our QB. In Sec. \ref{otpimization} we describe the VQE-inspired optimization for work extraction.  In Sec. \ref{results} we show our results. First, we use a specific example with only two qubits and then we explore the dependence of the work extraction with the number of qubits and the anisotropies of the Hamiltonian. Finally, in Sec. \ref{conclusion}, we conclude our work and discuss possible next steps. 

\section{Preliminaries and Methods}
\label{pre}
\subsection{Work extraction and ergotropy}
\label{Work extraction}

Given a closed quantum system described by the internal Hamiltonian $H$, the amount of extractable work by means of a reversible coherent operation is given by \cite{Alicki2013}
\begin{align}
	W(\rho,H)=E_{\rho}-{\rm Tr}\{H U \rho U^\dagger\}, \label{workextract}
\end{align}
where $U$ is a unitary operator and $E_{\rho}={\rm Tr}\{H\rho\}$ is the mean energy of the arbitrary state $\rho$. We assume the Hamiltonian with non-degenerate eigenvalues. In this scenario, the maximal work that can be extracted from the system is called {\it ergotropy} \cite{ergotropy2,ergotropy,Alicki2013,sone2021quantum} and it is obtained by minimizing the second term of Eq. (\ref{workextract}) over all possible unitary operations
\begin{align}
	\mathcal{E}\equiv W_{\rm max}(\rho,H)&=E_{\rho}-\min_U{\rm Tr}\{H U \rho U^\dagger\}\nonumber \\
	&=E_{\rho}-{\rm Tr}\{H\pi_\rho\},\label{ergotropy2}
\end{align}
where $\pi_\rho$ is called the passive state associated to $\rho$. By definition \cite{ergotropy2}, no work can be extracted from a passive state by means of unitary operations. For the case in which $\rho$ is a pure state, it is clear that the passive state is the ground state of the Hamiltonian $H$. However, for general mixed states, the passive state depends both on $H$ and the eigenvalues of $\rho$ \cite{ergotropy2}. We note from Eq. (\ref{ergotropy2}) that for reach the ergotropy it is assumed that any unitary operation can be performed on the system.  This assumption, however, is not usually true in realistic scenarios where one has a limited control over the systems. For this reason, to find protocols to optimize the work that can be extracted (or injected) from the system for different platforms is an interesting and current topic of research \cite{optimal,optimal2,optimal3,optimal4,optimal5,optent,VQErg,PhysRevLett.130.210401,PhysRevLett.120.117702}.

%The ergotropy has been extensively studied as the central figure of merit for extracting work from quantum systems, mainly in the context of quantum batteries \cite{Alicki2013,batteryphase,bibid}. Although several results for a multitude of different physical scenarios have been proposed in the last decade, they are mainly theoretical envisions and there is still a lack of experimental results dealing with work extraction \cite{bibid}. One of the major difficulties is to carry out the minimization in equation (\ref{workextract}), since the set of unitary operations that are available in practice are limited and depends on the specific experimental platform (i. e., the quantum hardware). 

\subsection{Models}
\label{spinchain}

In this work, we explore the amount of work that can be extracted from $N$-coupled spin-1/2 systems. To attain this task, we consider that the internal energy of the system is described by the following family of Hamiltonians ($\hbar=1$):
\begin{widetext}
	\begin{align}	
		H=-h\sum_{j=1}^N\sigma_z^{(j)}-J\sum_{j=1}^{N-1}[(1+\gamma)\sigma_x^{(j)}\sigma_x^{(j+1)}+(1-\gamma)\sigma_y^{(j)}\sigma_y^{(j+1)}+\Delta\sigma_z^{(j)}\sigma_z^{(j+1)}],\label{hamiltonians}
	\end{align}
\end{widetext}
where $\sigma_x^{(k)}$, $\sigma_y^{(k)}$, and $\sigma_z^{(k)}$ are the usual Pauli matrices for the $k$-th spin-1/2 system in the basis $\{\ket{\uparrow_k},\ket{\downarrow_k}\}$. For the sake of simplicity, from now on we are going to use the term qubits when referring to the spin-1/2 systems. The parameter $J$ is the coupling between the qubits while $\gamma$ and $\Delta$ are constants that introduce anisotropies in the chain.  The $h$ parameter is an external field which breaks the degeneracy of the energy levels of the Hamiltonian. By selecting specific parameters $\gamma$ and $\Delta$, the family of Hamiltonians (\ref{hamiltonians}) can be divided in the paradigmatic models described in Table \ref{model}.

\begin{table}[t]
	\centering
	\begin{tabular}{|l|l|}
		\hline
		{\bf \ Parameters} & {\bf \ Models}\\
		\hline\hline
		\ $\Delta=1$, $\gamma=0$ & \  $XXX$  \\ \hline
		\ $\Delta\neq0$, $\gamma=0$ & \  $XXZ$  \\ \hline
		\ $\Delta\neq0$, $-1<\gamma<1$ \ & \  $XYZ$    \\ \hline
		\ $\Delta=0$, $\gamma=0$ & \  $XX $   \\ \hline
		\ $\Delta=0$, $-1<\gamma<1$ \ & \  $XY$    \\ \hline
		\ $\Delta=0$, $\gamma=\pm1$ & \ Transverse Field Ising (TFI) \  \\ \hline 
	\end{tabular}
	\caption{Hamiltonian models.}	\label{model}
\end{table}

%\red{It is known that the hamiltonians (\ref{hamiltonians}) can have entangled eigenstates depending on the ratio $J/h$. This makes the optimization task even harder, since the passive state can be entangled. Then, to reach the ergotropy it is usually needed a set of entangling operations.}

\subsection{VQE-inspired optimization}
\label{otpimization}
\begin{figure*}[t]
	\centering
	\includegraphics[width=\textwidth]{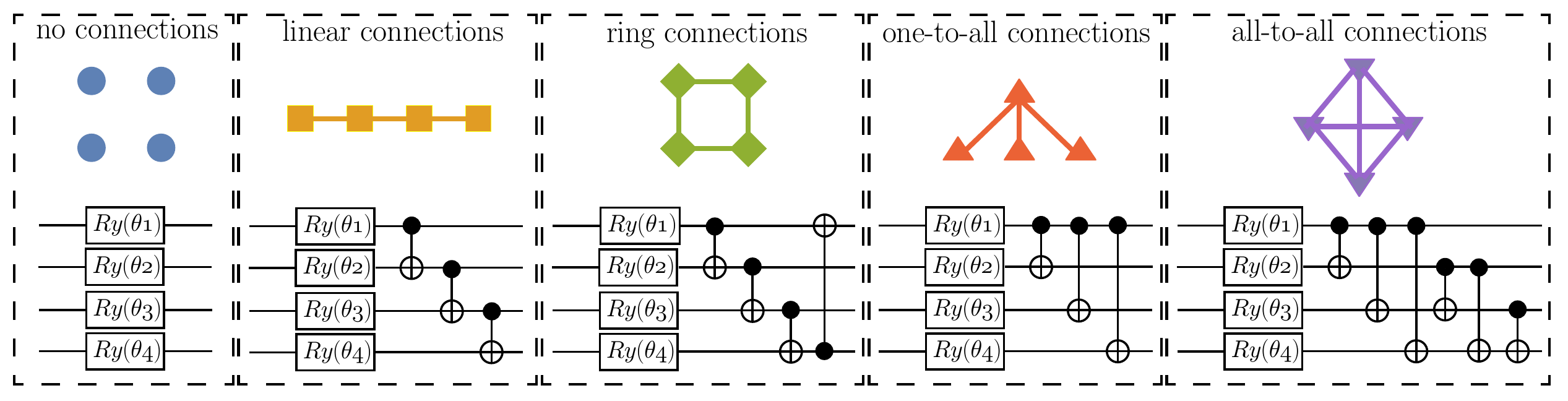}
	\caption{(colour online) Types of ansätze we are using in this work. The graphs represent the connectivity we are considering in each ansatz followed by the parameterized quantum circuit. The ansätze consist in a layer of local rotations followed by a specific type of connections, which in this work are built using CNOT gates. Those are examples for four qubits and can be directly generalized for $N$ qubits.}
	\label{ansatze}
\end{figure*}

The VQE optimization consists in two main ingredients. First, we need a parameterized circuit (ansatz) which imprints a set of parameters $\boldsymbol{\theta}=\{\theta_1,\theta_2,\cdots,\theta_j\}$ on the input state $\rho$. In general, the ansatz is expressed as a unitary parameterized operation $U(\boldsymbol{\theta})$. 
%In general, the ansatz is a unitary operation $U(\boldsymbol{\theta})$, that can be expressed as a product of successive $L$ unitaries \cite{Cerezo2021VqaReview}
%\begin{align}
%	U(\boldsymbol{\theta})=\mathcal{U}_1(\boldsymbol{\theta}_1)\phantom{.}\mathcal{U}_2(\boldsymbol{\theta}_2)\dots\mathcal{U}_L(\boldsymbol{\theta}_L),
%\end{align}
%with
%\begin{align}
%	\mathcal{U}_{\ell}(\boldsymbol{\theta}_{\ell})=\prod_{m}^{}e^{-i \theta_m \mathcal{O}_m}\Lambda_m.
%	\label{ul}
%\end{align}
%In Eq. (\ref{ul}), $\mathcal{O}_m$ is a hermitian operator and $\Lambda_m$ is an unparameterized unitary. 
In particular, we are going to use some ansätze belonging to the class of the hardware efficient ansatz (HEA) \cite{kandala2017hardware,Cerezo2021VqaReview,tilly2022variational}. The main characteristic of the HEA is that it can be tailored for a specific quantum device (experimental platform) in which one wants to run the protocol. In this class of ansätze, the unitaries are taken from a limited set of gates which are determined from the native connectivity for the chosen device. Moreover, in Ref. \cite{wu2021efficient} it was shown that HEA can be used to tailor more expressive parameterized operations using less entangling gates, which provides a better efficiency to the problem of finding ground states of some small molecules.  Adding to that, in Ref. \cite{BravoPrieto2020scalingof} it was shown that HEA can be used to find ground states of small spin chain systems with reasonable accuracy. As discussed in Ref. \cite{tilly2022variational}, using HEA in a VQE optimization process is a very convenient tool to prove principles when dealing with a small number of quantum systems. These later results motivate us even more using HEA in our work, since we focus on small controllable chains (up to eight qubits). For the purposes of this work, we can express the parameterized unitary $U(\boldsymbol{\theta})$ as
\begin{align}
U(\boldsymbol{\theta})=\prod_{k,j}\Lambda_{k,j}\prod_{m=1}^{N}R_y^{(m)}(\theta_m),
\label{ul1}
\end{align}
where $R_y^{(m)}(\theta_m)=e^{-i\theta_m \sigma_y^{(m)}}$ is a local rotation of the $m$-th qubit around the $y$-axis in the Bloch sphere, $\Lambda_{k,j}=\ket{\uparrow_k}\bra{\uparrow_k}\otimes\sigma_0^{(j)}+\ket{\downarrow_k}\bra{\downarrow_k}\otimes\sigma_x^{(j)}$ is a CNOT gate applied to the qubits $k$ (control) and $j$ (target) and $N$ denotes the number of qubits considered on the chain. Here, the operations $\Lambda_{k,j}$ express the connectivity of $U(\boldsymbol{\theta})$, i.e., the capability of the ansatz to generate entanglement between the qubits $k$ and $j$,  and the number of $\Lambda_{k,j}$ is specific for each ansatz. In our notation $\sigma_0^{(j)}$ is the identity operator for the qubit $j$. The ansätze we are considering in this work are depicted in Fig. \ref{ansatze}. As an example, we can cast the all-to-all (ata) connection ansatz for $N=4$ in the shape of Eq. (4) as 
\begin{align}
	U_{\rm ata}(\boldsymbol{\theta})=\left(\prod_{\substack{k=1, j=2 \\ k<j}}^{\substack{k=3, j=4}}\Lambda_{k,j}\right)^{\rm T}\prod_{m=1}^{4}R_y^{(m)}(\theta_m),
	\label{ul1}
\end{align}
where the transpose operation $(\cdot)^{\rm T}$ is taken to maintain the correct order of the CNOTs application, as shown in Fig. \ref{ansatze}. For the no connections (nc) ansatz only the local rotations $R_y^{(m)}(\theta_m)$ are performed.

%The ansätze we are considering in this work are shown in Fig. \ref{ansatze}, where $R_y(\theta_k)=e^{-i\theta_k \sigma_y^{(k)}}$ is a local rotation of the $k$-th qubit around the $y$ axis in Bloch sphere. Each depicted ansatz corresponds to a specific connectivity that may be available for a chosen quantum hardware.  
%In this class of ansätze, one uses unitaries that are taken from a limited set of gates determined from the connectivity and interactions that are specific to a quantum hardware, \st{which avoids} \blue{aimed to avoid} the circuit depth overhead arising from translating an arbitrary unitary into a sequence of gates easily implementable on a device. Here, we reduce our set of quantum gates to local rotations and CNOTs. The ansätze we are considering in this work are shown in Fig. \ref{ansatze}, where each one corresponds to a specific connectivity that can be available for a chosen quantum hardware.

The second ingredient is the definition of a cost (or loss) function which encodes our problem solution. As we are interested in the extractable work from the quantum source defined by the Hamiltonian in Eq. (\ref{hamiltonians}) and an input state $\rho$, we can naturally define the cost function as
\begin{align}
	W(\boldsymbol{\theta})=E_{\rho}-{\rm Tr}\{H U(\boldsymbol{\theta})\rho U^\dagger(\boldsymbol{\theta})\},\label{costf}
\end{align} 
where $W(\boldsymbol{\theta})$ is the extractable work in Eq. (\ref{workextract}) parameterized by the ansatz $U(\boldsymbol{\theta})$. Our task now is to optimize $W(\boldsymbol{\theta})$ for a given ansatz $U(\boldsymbol{\theta})$. The optimization is carried out using the following steps. We start with a random set of parameters $\boldsymbol{\theta}$ and register the value of $W(\boldsymbol{\theta})$. Then, we apply the gradient descent method, which consists in updating the set of parameters $\boldsymbol{\theta}$ using the iterative process
\begin{align}
	\theta_{j+1}\rightarrow\theta_j+k\frac{\partial W(\boldsymbol{\theta})}{\partial \theta_j},\label{iteration}
\end{align}
where $k$ is the step size taken towards the gradient. The iterative process in Eq. (\ref{iteration}) is repeated until the cost function converges. Then, we obtain a set of parameters leading to $W(\boldsymbol{\theta}_{\rm opt})\equiv \mathcal{W}$, where $\mathcal{W}$ is the optimal amount of work that we can extract from $\rho$ using a given ansatz described by a parameterized circuit.
\begin{figure*}[t!]
	\centering
	\includegraphics[width=\textwidth]{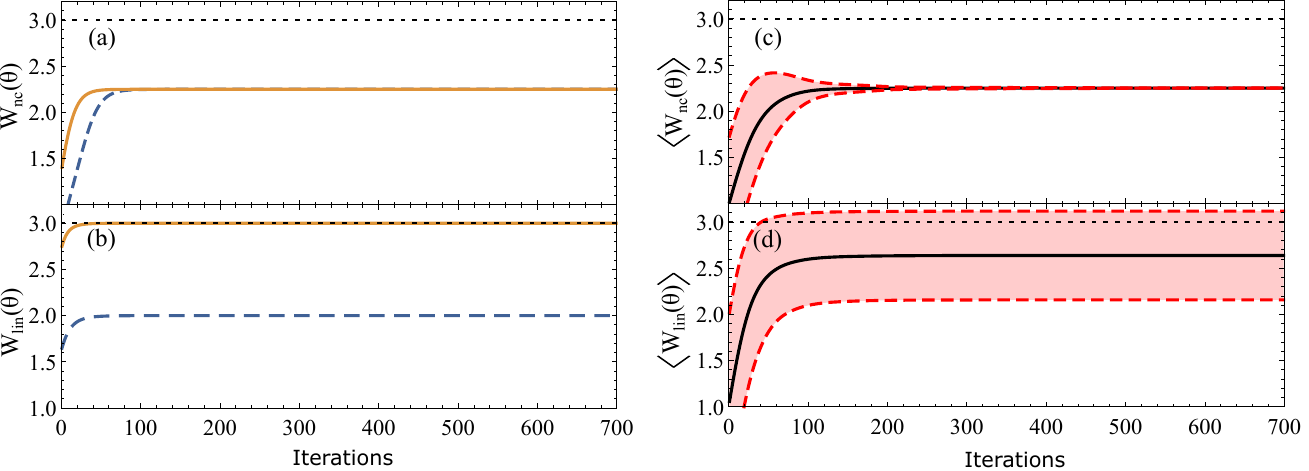}
	\caption{Panels (a) and (b) show the convergence of the cost functions given by Eqs. (\ref{nc}) and (\ref{lin}) for two different sets of initial random parameters $\{\theta_1,\theta_2\}$. Panels (c) and (d) show the convergence of Eqs.(\ref{nc}) and (\ref{lin}) on average after repeating the optimization process 2000 times with initial parameters chosen randomly in the interval $[0,\pi]$. The red dashed lines depict the standard deviation from the average.}
	\label{convergence}
\end{figure*}

\section{Results}
\label{results}
In what follows, we are going to show our results for different number of qubits in the chain described by the family of Hamiltonians in Eq. (\ref{hamiltonians}). For all simulations in this work, we are going to fix the input state as (unless otherwise stated) 
\begin{align}
\rho=\bigotimes_{j=1}^{N}\ket{\uparrow_j}\bra{\uparrow_j}.\label{input}
\end{align}
This state is a fair assumption to illustrate our results since it can be prepared in the most of experimental platforms by means of projective measurements and other state initialization techniques. We can mention, for instance,  super conducting qubits hardware in IBM quantum computers \cite{ibm1,ibm2}, where these initial states, Eq. (\ref{input}), and most of the ansätze could be implemented. In trapped ions quantum hardware \cite{Iontraps}, it is also possible to initialize the system in these states and implement the all-to-all connections, which is the most complex one [see Fig. \ref{ansatze}]. Another platform where initialization and control techniques to attain the tasks proposed can be reached is nuclear magnetic resonance (NMR) \cite{expMahesh,NMRvqa}. We also fix the couplings $J=-1.0$ a.u. and $h=1/2$ a.u., where a.u. holds for arbitrary units. All of the following simulations were done using the software {Mathematica$^\text{\textregistered}$ with help of the Melt library \cite{melt}. 
\begin{figure*}[t]
	\centering
	\includegraphics[width=\textwidth]{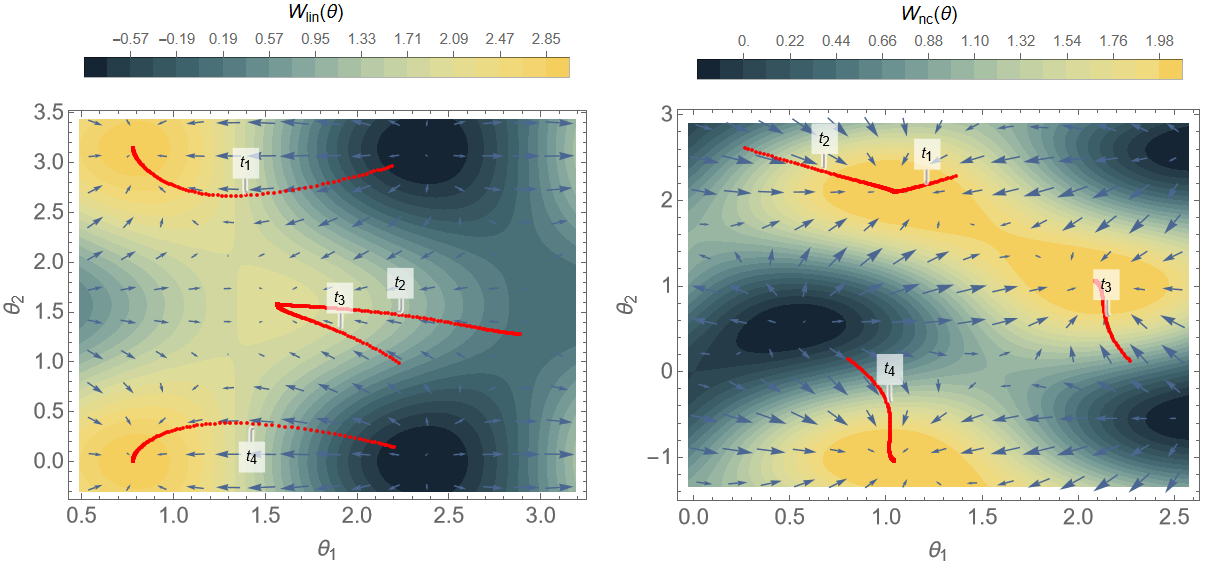}
	\caption{Plots of the cost functions landscapes. The red dots track the optimization trajectory $t_i$ of the parameters $\theta_1$ and $\theta_2$. For each $t_i$ we use a different set of random initial parameters. The functions in Eqs. (\ref{nc}) and (\ref{lin}) are periodic presenting multiple global maxima, as we can see from the landscapes. }
	\label{2qlandscapes}
\end{figure*}

\subsection{Work extraction from two interacting qubits}

We start by showing our results for only two qubits ($N=2$). This case is illustrative because there are only two parameters to optimize $\{\theta_1,\theta_2\}$. Then, the cost function can be visualized as a three-dimensional (3D) function (cost function landscape), from which we can grasp some of the main features of the VQE-optimization process for work extraction. In this spirit, we choose the $XX$ model as an example
\begin{align}
H=-\frac{1}{2}\left(\sigma_z^{(1)}+\sigma_z^{(2)}\right)+\left(\sigma_x^{(1)}\sigma_x^{(2)}+\sigma_y^{(1)}\sigma_y^{(2)}\right) \label{XX2},
\end{align}
where the external field parameter is $h=1/2$ a.u.
This Hamiltonian can be diagonalized, from which we can find its eigenstates
\begin{align}
&\ket{E_0}=\frac{1}{\sqrt{2}}\left(\ket{\downarrow\uparrow}-\ket{\uparrow\downarrow}\right)\label{singlet},\\
&\ket{E_1}=\ket{\downarrow \downarrow},\\
&\ket{E_2}=\ket{\uparrow \uparrow},\\
&\ket{E_3}=\frac{1}{\sqrt{2}}\left(\ket{\downarrow\uparrow}+\ket{\uparrow\downarrow}\right),
\end{align}
with  associated energies $\{E_0,E_1,E_2,E_3\}=\{-2,-1,1,2\}$ a.u.. As we can see, this model has a Hamiltonian spectrum with entangled eigenstates. In particular, the ground state is the singlet state, Eq. (\ref{singlet}). For two qubits, there are only two ansätze available for our protocol: linear connection between the qubits and no connections (local rotations only). For both cases, the cost function (\ref{costf}) can be straightforwardly computed and their analytical expressions are given by
\begin{widetext}
\begin{align}
	&W_{\rm nc}(\boldsymbol{\theta})=E_{\rho}-\cos^2(\theta_1)\cos^2(\theta_2)+\sin^2(\theta_1)\sin^2(\theta_2)-4\cos(\theta_1)\cos(\theta_2)\sin(\theta_1)\sin(\theta_2),\label{nc}\\
	&W_{\rm lin}(\boldsymbol{\theta})=E_{\rho}-\left[4\cos(\theta_1)\sin(\theta_1)\cos(2\theta_2)+\cos^2(\theta_1)\sin^2(\theta_2)-\sin^2(\theta_1)\cos^2(\theta_2)\right],\label{lin}
\end{align}
\end{widetext}
where the subscripts are for the no connections case (nc) and for the linear connection (lin) (see Fig. \ref{ansatze}). The input state is $\rho=\ketbra{E_2}{E_2}$ which has a mean energy of $E_\rho=1.0$ a.u. In this case, the passive state is $\pi_\rho=\ketbra{E_0}{E_0}$, then the  ergotropy [see Eq. (\ref{ergotropy2})] is readily given by $\mathcal{E}=3.0$ a.u..

First, let us discuss about the convergence for the different cost functions given by Eqs. (\ref{nc}) and (\ref{lin}) when running the optimization process. In Fig. \ref{convergence} we show the plots for Eqs. (\ref{nc}) and (\ref{lin}) as a function of the iteration steps. We start by looking at the performance of the ansatz with no connections between the qubits. In Fig. 2(a) it is shown two examples of convergence for $W_{\rm nc}(\boldsymbol{\theta})$, each one is initialized with a different set of  initial parameters $\{\theta_1,\theta_2\}$. The initial parameters are randomly selected in the interval $[0,\pi]$. It is clear that both examples converge for the same value $\mathcal{W}_{\rm nc}\approx2.3$ a.u., which is far from the ergotropy value of $3.0$ a.u. (shown as the black dotted line). However, in Fig. 2(b) we see that for different initial parameters, the cost function $W_{\rm lin}(\boldsymbol{\theta})$ does not always converge to the same value. While the trajectory described by the yellow solid line reaches the ergotropy, the blue dashed line reaches a value of $\mathcal{W}_{\rm lin}\approx2.0$ a.u., which is less than the one obtained with the no connections ansatz. Taking this into account, it is reasonable to look at the convergence of the cost functions on average. To achieve this, we repeat the optimization process $M$ times  to obtain the average optimal extractable work $\langle\mathcal{W}\rangle$. In Figs. 2(c) and 2(d) we show the convergence of the cost functions averaged over $M=2000$ realizations of the optimization process. We can see that it is possible to extract more work on average using entangling operations, i.e., $\langle\mathcal{W}\rangle_{\rm lin}>\langle\mathcal{W}\rangle_{\rm nc}$. This is somehow expected, since the passive state is an entangled state and the input state [see Eq. (\ref{input})] is separable. The same plots show the standard deviation (red dashed line) as a function of each iteration step. It is interesting to remark that, although we can extract more work on average using the linear ansatz, its dispersion is larger than the ansatz with no connections, which does not have any dispersion after the convergence.

%where $\langle \mathcal{W} \rangle$ optimal work that can be extracted in average repeating optimization process $M$ times with randomized initial parameters. In our simulations, we use $M=2000$ to compute the average optimal work.  In plots (b) and (e) it is shown $\eta_{\rm nc}$ and $\eta_{\rm lin}$ (black solid lines), respectively. 

To deepen our understanding about the average behavior of the cost functions, in Fig. \ref{2qlandscapes} we show  the cost function landscape for both ansätze. The vectors within the contour plots represent the gradient of the cost function. The red lines track some trajectories of the parameters in the parameter space during the optimization process. Each trajectory is obtained from a different set of initial random parameters. For the ansatz with no connections we see that even though we have different trajectories, they always converge to the global maxima of the cost function. This is not the case for the linear connection ansatz. While the trajectories $t_1$ and $t_4$ converge to the maximum value of the cost function, the trajectories $t_2$ and $t_3$ get stuck in a region of the parameter space where the gradient is infinitely small (the norm of the gradient vector is represented by the arrow size). In this case, this region is due to the occurrence of local maxima in the cost function. However, as the total dimension increases, it is known that the HEA can lead to regions with highly dense local maxima and barren plateaus \cite{CerezoAkira2021BarrenPlateaus,cerezo2022diagnosing,liu2022laziness,martin2023barren,mcclean2018barren,holmes2022connecting,tilly2022variational}, which are large regions in the cost function landscape of parameters where the gradient vanishes. It is also worth remarking on the fact that the functions given by Eqs. (\ref{nc}) and (\ref{lin}) have a period $\pi$. Then, these functions present multiple global maxima by periodicity. %as those reached by the trajectories $t_1$, $t_2$ and $t_4$ in the linear cost function landscape shown in Fig. 3, for example. }

\subsection{Work extraction efficiency from $N$ interacting qubits}
\begin{figure*}[t]
	\includegraphics[width=\textwidth]{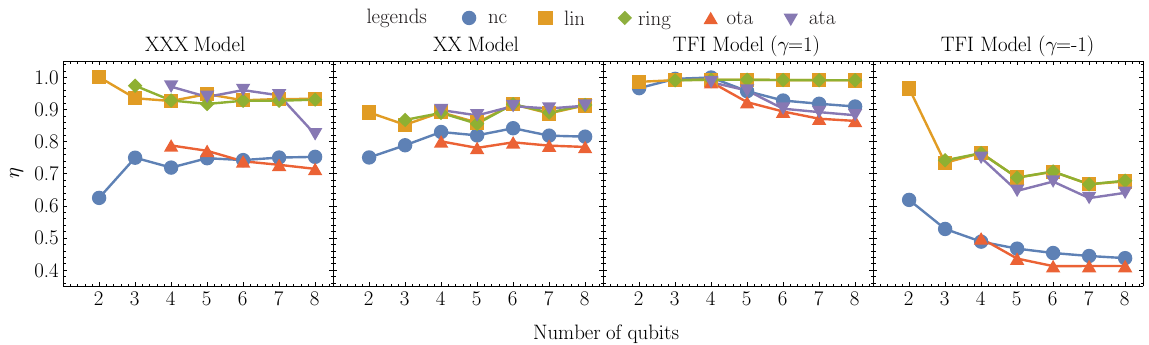}
	\caption{(colour online) Efficiency $\eta$ as in Eq. (\ref{effi}) as a function of $N$ for the $XXX$, $XX$ and TFI models. Each pattern corresponds to a different ansatz as depicted in Fig. \ref{ansatze}. The plot legends on the top means: no connections (nc), linear connections (lin), ring connections (ring), one-to-all connections (ota), and all-to-all connections (ata). }
	\label{symmmod}
\end{figure*}

We shall now present our results for chains up to eight qubits. It is convenient to define the efficiency of the work extraction protocol as
\begin{align}
	\eta=\frac{\langle \mathcal{W}  \rangle}{\mathcal{E}},\label{effi}
\end{align} 
which is a dimensionless quantity varying from 0 to 1. This efficiency quantifies how close we can get to the ergotropy using a given ansatz. First, we analyze the performance of our protocol for the paradigmatic $XXX$, $XX$ and TFI models. In Fig. \ref{symmmod}, we show the behavior of $\eta$ with the number of qubits for each model using the ansätze described in Fig. \ref{ansatze}. An interesting result is that the ansatz with no connections, i.e., no entangling operations, achieves a high efficiency (more than 0.9) for the TFI model with $\gamma=1.0$ a.u.. Specifically, for three and four qubits, an efficiency about $\eta\approx0.99$ is achieved. Another interesting result, is that the efficiency obtained using the one-to-all connections ansatz decreases as the number of qubits in the chain increases. For more than five qubits, the efficiency of this ansatz is the worst one for all shown cases. This may be explained by the fact that the connectivity of the one-to-all ansatz is incompatible with the structure of the Hamiltonian, which has couplings between first neighbors only. Then, one does not expect the ground state (which is the passive state) to have entanglement between distant qubits in the chain. This also explains why the linear, ring, and all-to-all connection ansätze, which have entangling operations between first neighbors, presented the best efficiencies for all models. We stress here that the all-to-all ansatz produce both, long-range and first neighbor correlations, in contrast with the one-to-all ansatz, which only generates long range correlations.

Now, let us turn our attention to the TFI model with $\gamma=-1.0$ a.u.. As we can see, its efficiency decays as $N$ increases for all ansätze, differently from what happens for the TFI model with $\gamma=1.0$ a.u.. This suggests that the anisotropy parameter $\gamma$ plays an important role in the amount of work that can be extracted given a specific ansatz. Taking this into account, we analyze the role of the anisotropy by looking at the efficiency as a function of $\gamma$ for the models $XY$ and $XYZ$, as shown in Fig. \ref{xyzmod}. For the $XY$ model, we clearly see that the amount of work that can be extracted from the spin chain increases as the anisotropy parameter $\gamma$ varies from $-1.0$ a.u. to $1.0$ a.u. for all ansätze used. The $XYZ$ model exhibits a similar result for almost all ansätze. For $N=8$, we see that the efficiency obtained with the all-to-all ansatz suddenly decreases when compared with its value obtained for fewer qubits. 
%This may be explained by the fact that this ansatz is very expressive \textcolor{red}{teria que calcular a expressividade pra confirmar isso, não me parece uma afirmação justa} \cite{}, and as the number qubits increases it can lead to Barren Plateau problems \cite{}. 
%\textcolor{red}{acho melhor tirar o "due to overparameterization" pq aparentemente isso ajuda a escapar de minimos locais, não o contrário - https://arxiv.org/pdf/2205.05786.pdf, https://www.nature.com/articles/s41534-023-00681-0.pdf, https://arxiv.org/pdf/2307.03292.pdf... as vezes a razão é por ele ser mais expressível ou por ele nao bater com o hamiltoniano mesmo}

\begin{figure*}[t]
	\includegraphics[width=\textwidth]{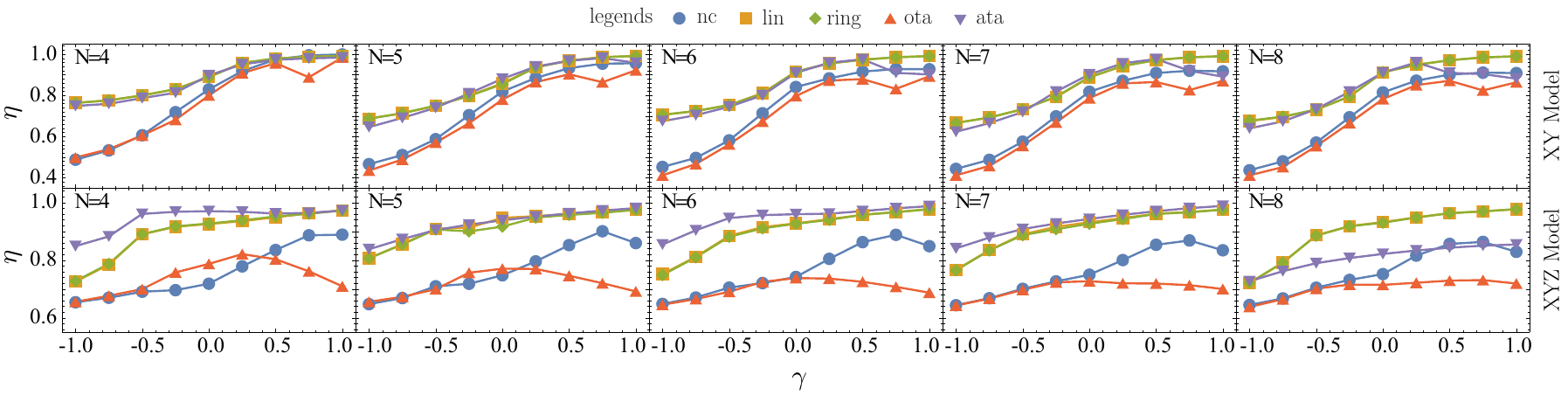}
	\caption{(colour online) Efficiency $\eta$ as in Eq. (\ref{effi}) as a function of the anisotropy parameter $\gamma$ for the $XY$ (top panel) and $XYZ$ (bottom panel) models with different number of qubits. The anisotropy parameter $\gamma$ is varied from $-1.0$ a.u. to $1.0$ a.u. in steps of $0.25$ a.u. For the XYZ model we fixed $\Delta=1.0$ a.u. for all plots. Each pattern corresponds to a different ansatz as depicted in Fig. \ref{ansatze}.  The plot legends on the top means: no connections (nc), linear connections (lin), ring connections (ring), one-to-all connections (ota), and all-to-all connections (ata).}
	\label{xyzmod}
\end{figure*}
\begin{figure*}[t]
	\includegraphics[width=\textwidth]{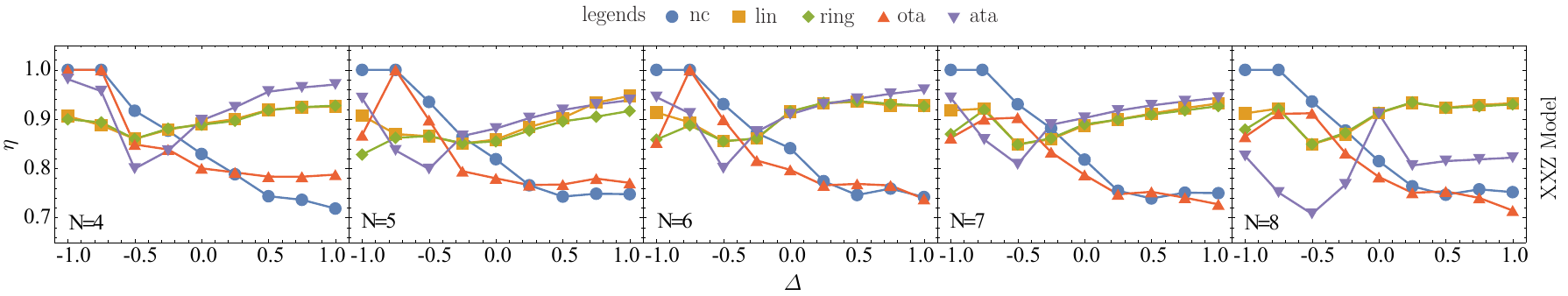}
	\caption{(colour online) Efficiency $\eta$ as in Eq.(\ref{effi}) as a function of the parameter $\Delta$ for the $XXZ$ model. The parameter $\Delta$ is varied from $-1.0$ a.u. to $1.0$ a.u. in steps of $0.25$ a.u. Each pattern corresponds to a different ansatz as depicted in Fig. \ref{ansatze}.  The plot legends on the top means: no connections (nc), linear connections (lin), ring connections (ring), one-to-all connections (ota), and all-to-all connections (ata).}
	\label{xxzmod}
\end{figure*}

Finally, let us discuss about how the work extraction efficiency behaves as we vary the parameter $\Delta$ for the $XXZ$ model. In Fig. \ref{xxzmod}, it is shown the plots for the efficiency as a function of $\Delta$  for the ansätze in Fig. \ref{ansatze} up to eight qubits. From the plots, we can see an interesting behavior. For $\Delta<-0.25$ a.u. the no connections ansatz is the best option, and is able to extract the ergotropy for $\Delta=\{-1.0,-0.75\}$ a.u. For $\Delta>-0.75$ a.u., the entangling ansätze (with the exception of the one-to-all ansatz) become more efficient than the no connections ansatz. This suggests  a change on the passive state structure, which probably changes from a separable state to an entangled one.

\section{Conclusions}
  
In this work we proposed a VQE-inspired optimization to investigate the amount of work that can be extracted from first neighbor coupled qubits. We compared the efficiency of work extraction assuming different types of connectivity. Our simulation results show that the ansätze that generate entanglement between first neighbors (i.e., linear, ring and all-to-all connections) have the best performance while the one-to-all ansatz has, in general, the worst. This can be explained by the fact that the connectivity of the one-to-all ansatz is not compatible with the structure of first neighbor Hamiltonians. %(\red{tem um paper que fala disso mas não to achando})
Our results also suggest that using only local rotations can be a good strategy depending on the connectivity available and the setup of the QB (Hamiltonian and number of qubits). It is worthwhile to remark that, although the all-to-all present the best results, the number of CNOTs in this ansatz scales with $N(N-1)/2$, while for the linear and ring ansätze it scales with $N-1$ and $N$, respectively. In practice, the all-to-all ansatz can be problematic depending on the number of qubits in the QB since CNOT gates are more prone to errors when compared to local operations.

A natural step for future works would be explore how long-range interactions in the qubits chain (as, e.g., considered in Ref. \cite{Tao}) would affect the efficiency of the VQE-inspired optimization. In these cases, it is not obvious  we will have the same hierarchy behavior, where the ansätze that have couplings between first neighbors have the best performances. Considering the effects of disordered spins also would be interesting. One could also explore how the efficiency of work extraction would be impacted by increasing the number of layers used in the ansätze shown in Fig. \ref{ansatze}. In Ref. \cite{BravoPrieto2020scalingof}, it was demonstrated that the efficiency in finding ground states of some spin chains can be improved by considering a number of layers at least linear in the system size when using HEA-type structures built with local rotations and CZ gates. Connected to this latter work,  our protocol can naturally find applications for the problem of preparing/finding the ground state of many-body Hamiltonians. At this point, however, it is worth remarking that there are limitations when using the HEA class to solve optimization tasks. Even though it is a good choice for proof of principles in small systems as we stated before, for more complex and large systems such as molecular ones, the HEA is not expected to be a good option \cite{tilly2022variational}. One of the issues, is that HEA is very inefficient to span very large Hilbert spaces, requiring circuits that may have exponential depth. Due to that, the HEA is very prone to  barren plateau problems \cite{CerezoAkira2021BarrenPlateaus,cerezo2022diagnosing,liu2022laziness,martin2023barren,mcclean2018barren,holmes2022connecting,tilly2022variational}. Nevertheless, finding optimal HEA structures and develop optimal ansätze targeted to specific complex tasks are both ongoing topics of research. 

It is also important to remark that in this work we fixed the initial state as Eq. (\ref{input}). A very exciting next step would be exploring how the VQE-inspired optimization protocol performs for different classes of initial states. In tasks such as energy extraction from QBs or thermal machines, the state from which energy has to be extracted is dependent on the specific charging protocol (for QBs) or machine cycle. Moreover, the preparation of the initial state to obtain better results when optimizing a certain task using some VQA is a currently topic of research which, to the best of our knowledge, has not been universally addressed. Furthermore, the amount of resources to prepare arbitrary states as the number of qubits grown it is not a trivial issue to address. For instance, an interesting discussion about the electrical energy cost of arbitrary state preparation in the context of photonic integrated circuits is provided in Ref. \cite{pierre}.  With the protocol we presented in this work, we expect to shed some light in these tasks, since it can be applied to any initial state when treating a specific protocol. 
%\red{There are however many known limitations of HEA. A first obvious issue is the fact that it must span a very large
%	portion of the Hilbert space to guarantee that an accurate enough representation of the ground state wavefunction can
%	be produced. HEA can therefore be quite inefficient with respect to its required Hilbert space coverage, requiring in
%	worst cases an exponential depth (though further research would be required to verify how the accuracy of ground state
%	representations would respond to arbitrary depth reduction and shrinkage of the Hilbert space coverage). An immediate
%	consequence is that HEA is significantly limited by barren plateaus [127] (see Section 6.1 [86]). It is, however, unlikely to
%	be suitable for general larger-scale chemical problems. One therefore would rather turn toward more problem-tailored
%	ansätze.} 

Finally, we would like to mention that recently a variational algorithm to estimate the ergotropy from quantum many-body batteries was proposed in Ref. \cite{VQErg}. In this work the authors developed an algorithm, called variational quantum ergotropy (VQErgo), to find the optimal unitary for the work extraction from the QB. In their work, the QB is modelled by the internal local Hamiltonian $H=-h\sum_j \sigma_z^{(j)}$ and subjected to a drive which induces a tranverse Ising coupling used to charge or discharge the QB. Although similar in spirit, it is a very different approach from ours. First, we considered our QB to be encoded in a spin chain Hamiltonian. Second, our ansätze were built taking into account the connectivity that may be available for different quantum platforms. That being said , we truly believe that both approaches give important contributions to the areas of quantum thermodynamics and variational quantum algorithms.

\label{conclusion}

\begin{acknowledgments}
\noindent 
The authors thank professors Sérgio Muniz and Askery Canabarro for the fruitful discussions.
I.M. acknowledges financial support from São Paulo Research Foundation - FAPESP (Grant No. 2022/08786-2). This study was financed in part by the Coordenação de Aperfeiçoamento de Pessoal de Nível Superior – Brazil (CAPES) – Finance Code 001. P.C.A. acknowledges financial support from Conselho Nacional de Desenvolvimento Científico e Tecnológico - Brazil (CNPq - Grant No. 160851/2021-1). D.O.S.P acknowledges the support by the Brazilian funding agencies CNPq (Grant No. 304891/2022-3), FAPESP (Grant No. 2017/03727-0) and the Brazilian National Institute of Science and Technology of Quantum Information (INCT/IQ). 
\end{acknowledgments}

\bibliographystyle{unsrt}
\bibliography{allrefs.bib}

\end{document}